\documentclass[12pt]{article}
\usepackage{tikz}
\usepackage{amsmath}
\usepackage{mathtools}
\usepackage{fancyhdr}
\usepackage{gensymb}
\usepackage{graphicx}
\usepackage{cite}
\usetikzlibrary{decorations.pathmorphing, arrows, patterns}
\tikzset{snake it/.style={-stealth,
decoration={snake, 
    amplitude = .4mm,
    segment length = 2mm,
    post length=0.9mm},decorate}}
\usepackage{amssymb}
\usepackage{endnotes}
\let\footnote\endnote

\usepackage[a4paper, total={6in, 10in}]{geometry}

\begin{document}
\title{\Large{\textbf{On the symmetry of a one-dimensional hydrogen atom}}}
\date{}
\author{Boris Iveti\'c
 \\ Department of Physics, Faculty of Science, University of Zagreb
  \\ Bijeni\v cka cesta 32, 10000 Zagreb, Croatia \\}

\maketitle

\begin{abstract}
We touch upon a long-standing question of the "true" one-dimensional hydrogen atom solution. From a symmetry point of view, Kepler problem in $d\ge2$ dimension is characterized by geometrical rotational symmetry, $SO(d)$, as well as dynamical, "accidental" $SO(d+1)$ symmetry. Because of topology, these two symmetries are mutually exclusive in one dimension, regardless of the regularization employed, drawing one to a conclusion  that the question of "true" hydrogen atom in one dimension doesn't have an answer because a single dimension can not support both of the symmetries of Kepler  problem. We argue our findings using a novel method to recover and classify solutions appearing in the literature according to the symmetry they respect. In particular, curious features of some of the solutions - double degeneracy and particle confinement - are directly attributed to the dynamical symmetry behind them. 
\end{abstract}

\section{Introduction}
The simplest example of a potential containing nonintegrable singularity is the 1d
$-e^2 /|x|$ potential. This quantum mechanical "nuance" has a half-century long history
of scientific research\cite{lou, schw, and, hai, ges, imb, ny, dav, mos1, new, mos2, fis, fis2, kur, gor, pr}.
While solutions away from the origin are easy enough to find \cite{lou}, there is no
unique way of patching them up at the point of singularity. In fact, there exists a
four-parameter family of self-adjoint extensions of the Hamiltonian \cite{fis}, corresponding
to the different boundary conditions on the wave-function and it’s derivative at the
origin. However, mathematics alone cannot determine which of the possible self-
adjoint extensions corresponds to a particular physical problem, meaning that one
needs additional physical input for the boundary conditions at the origin to uniquely
define the problem.

Indefiniteness of the Hamiltonian can be traced to indefiniteness of the $x\to 0$
limit of the potential. This was argued to be Dirac's delta function, or some combination of Dirac's deltas and their derivatives \cite{gor}, giving different boundary
conditions on the solutions at the origin. For instance, in \cite{gor} it was argued that this
limit should be Dirac's delta function in order to have zero classical force at the
equilibrium position $x = 0$. Alternatively, the arbitrariness was also interpreted as
coming from different zero-coupling limits of the potential \cite{fis}. As we will show below, this arbitrariness can also be ascribed to the different ways of reducing higher
dimensional problem to a single dimension.

Specifically, two interesting and important questions arose in this context: the question of the double
degeneracy of energy levels, obtained by some authors, with respect to 1d no-degeneracy
theorem\cite{lou}, and the question of the $x = 0$ "barrier" penetrability\cite{mos1, new, mos2} 
\footnote{In some of the early papers there appeared yet another problem, that of the infinite energy ground-state solution, which required removal\cite{and, ges}}. Different approaches resulted in an interesting scientific discussion over the decades,
with a rich variety of methods and lots of interesting interplay between mathematics and physics. For instance, put forward were various regularization prescriptions for "smoothing
out" the singularity at the origin \cite{lou, hai}; in \cite{ny} the
problem was solved in the momentum space defining $1/|x|$ as an integral operator; \cite{gor} used a Laplace transform approach,
with a detailed classical and semi-classical analysis, while in \cite{fis, fis2} functional-integral and
functional-analytic approaches were used. \footnote{For the reasons of brevity, we do injustice to everyone whose contribution we don't mention here: for a very detailed analysis of chronological development of the problematic, we refer an interested reader to \cite{pr}; (s)he may also find useful an extensive references list of \cite{fis}.} However, we report here our
impression that a definite consensus on the number of physical solutions to
this problem has not been reached yet.

 In this paper we analyse symmetry of the problem, motivated in particular by the fact that this important aspect of the Kepler problem has been  almost completely ignored.\footnote{A nice pedagogical exposition of the above mentioned and other symmetries of a (3d) hydrogen atom is given in chapter 14 of \cite{gil}.}  We start by introducing a novel method using a regularized Fourier transform to pass to momentum space, which is then interpreted as a stereographic projection
from a circle, in analogy with $d\ge2$ dimensional Kepler problem\cite{ali}. Pushing on the
symmetry analyses, it is shown that the solutions can be either
parity or $SO(2)$ symmetric, but can not be both, regardless of the regularization procedure. Interpreting parity as a (remnant of) geometric rotational symmetry of Kepler problem when dimension goes to one, we conclude that at least one of the symmetries of the Kepler problem gets broken when trying to formulate it in one dimension. In the following we shall at several places refer to parity as "$SO(1)$" to emphasize this connection.
\footnote{We counted only three papers analysing symmetry aspects: \cite{dav} reproduces and recognizes $SO(2)$ solutions; \cite{gor} argues against this symmetry and advocates parity odd solutions, while \cite{ny} reproduces without recognition $SO(2)$ symmetry solutions and recognizes "spontaneous parity symmetry breaking", though connection with $SO(1)$  is not made.}

At the end of this introduction, we point out that the problem of a one-dimensional
hydrogen atom is today no more of purely academic interest (in fact it never was:
even the first appearance of the problem \cite{lou} was motivated by a model of exciton in high magnetic field)
- the number of physical systems that essentially exhibit 1d behaviour (nanowires,
trapped ions in laser created 1d potential, 1d systems in materials science, to name
only a few), with interactions described by an effective Coulomb-like potential, is only
growing.

\section{The Fourier transformed Schroedinger equation}
The Schroedinger equation of the 1d hydrogen atom is
\begin{equation}\label{coord}
-\frac{d^2}{dx^2}\psi(x)-\frac{2me^2}{|x|}\psi(x)=2mE\psi(x).
\end{equation}

 To apply the method of Fock, a first step would be to Fourier transform this equation. Here one encounters the problem of the Fourier transform 
 of  $1/|x|$, which does not exist  in one dimension, as opposed to higher dimensions. 

 Written explicitly, it is
\begin{equation}
\mathcal{F}\left(\frac{1}{|x|}\right)= \sqrt{\frac{2}{\pi}}\int_0^\infty\frac{\cos(px)}{x}dx,
\end{equation}
with the above integral logarithmically divergent due to the singularity at the origin.  
Noting that the cosine integral near the origin can be represented as 

\begin{equation}
\int_\epsilon^\infty\frac{\cos(px)}{x}dx=-\log|\epsilon p|-\gamma-\sum_{k=1}^\infty\frac{\left(-(\epsilon p)^2\right)^k}{2k(2k)!},
\end{equation}
where $\gamma$ is Euler-Mascheroni constant, we propose to first formulate the problem on a coordinate space domain $\mathbf{R}\backslash\langle-\epsilon_L,\epsilon_R\rangle$
with arbitrary positive $\epsilon_L$ and $\epsilon_R$. All the calculation can be formally done with  finite $\epsilon$'s, and the limit in which they go to zero can be explored at the very end. The purpose of this approach is to draw generic conclusions about the symmetry of the problem without the need to ever explicitly define this limit.

To have simpler expressions, we consider $\epsilon_L=\epsilon_R\equiv \epsilon$ throughout the derivation, with all final results straightforwardly generalizable to asymmetric case.

We write
\begin{equation}\label{ftpot}
\mathcal{F}\left(\frac{1}{|x|}\right)_\epsilon =\sqrt{\frac{2}{\pi}}\int_\epsilon^\infty\frac{\cos(px)}{x}dx =-\sqrt{\frac{2}{\pi}}\left(\log|\epsilon p|+\gamma+S(\epsilon p)\right)
\end{equation}
where  
\begin{equation}
S(\epsilon p)\equiv\sum_{k=1}^\infty\frac{\left(-(\epsilon p)^2\right)^k}{2k(2k)!}.
\end{equation}
If a Fourier transform of the wave function is

\begin{equation}
\mathcal{F}(\psi(x))=\frac{1}{\sqrt{2\pi}}\int^{\infty}_{-\infty}\psi(x)e^{ipx}dx\equiv
\phi(p),
\end{equation}
then convolution theorem gives

\begin{equation}\label{konv}
\mathcal{F}\left(-\frac{1}{|x'|}\psi(x')\right)_\epsilon=\frac{1}{\pi}\int_{-\infty}^\infty dp'\lbrace\log|\epsilon p'-\epsilon p|+\gamma+S(\epsilon p'-\epsilon p)\rbrace\phi(p').
\end{equation}
Notice that the validity of the Fourier transform in the limiting case $\epsilon \to0$ depends on the proper definition of the above expression, and not (\ref{ftpot}), for which this limit never exists. 

Now, the regularized Fourier transform of equation (\ref{coord}) is

\begin{equation}\label{blah}
(p^2+p_0^2)\phi(p)=-\frac{2me^2}{\pi}\int_{-\infty}^\infty dp'\lbrace\log|p'-p|+\log\epsilon+\gamma+S(\epsilon p'-\epsilon p)\rbrace\phi(p'),
\end{equation}
where we introduced a positive constant $p_0\equiv\sqrt{-2mE}$.
 
 The benefit of working in a momentum representation is already apparent from the above expression. Firstly, regularization dependent and regularization independent parts separate nicely. This allows us to study them separately. Secondly, a kernel of the integral eigenvalue operator is evidently symmetric, regardless of the regularization when $\epsilon\to0$ \footnote{In agreement with Kurasov \cite{kur}, who showed symmetry of Hamiltonian (in coordinate space).}. 
 \section{Fock's method in 1D}
Following the idea of Fock, we project the momentum space   from $\mathbf{R^1}$ to $\mathbf{S^1}$. We define an angle $\alpha$ by
\begin{equation}
\sin\alpha=\frac{2p_0p}{p_0^2+p^2} \ \ \ \ \ \ \ \ \cos\alpha=\frac{p_0^2-p^2}{p_0^2+p^2},
\end{equation}
with an inverse relation
$$p=p_0\tan\frac{\alpha}{2},$$
and with the range of $\alpha$ from $-\pi \ (p\to -\infty)$ to $\pi \ (p\to\infty)$. 
The infinitesimal elements are related by
$$dp=\frac{p_0}{2}\cos^{-2}\frac{\alpha}{2}d\alpha=\frac{p_0^2+p^2}{2p_0}d\alpha.$$
Defining a new wave function in terms of the variable $\alpha$ as  
\begin{equation}
\chi(\alpha)\equiv\frac{p_0^2+p^2}{p_0^2}\phi(p),
\end{equation}
with factor in front defining normalization condition
$$\frac{1}{2\pi}\int_{-\pi}^\pi d\alpha|\chi(\alpha)|^2=\frac{1}{\pi p_0}\int_{-\infty}^\infty dp\frac{p^2+p_0^2}{p_0^2}|\phi(p)|^2=\frac{2}{\pi p_0}\int_{-\infty}^\infty dp|\phi(p)|^2=\frac{2}{\pi p_0}\int_{-\infty}^{\infty} dx|\psi(x)|^2=1,$$
we write the projection of (\ref{blah}) onto a circle:
\begin{equation}
\begin{split}
\chi(\alpha)=-\frac{me^2}{\pi p_0}\int_{-\pi}^{\pi}d\alpha'\chi(\alpha') &
\lbrace\log\left|\tan\frac{\alpha}{2}-\tan\frac{\alpha'}{2}\right|+\log|p_0\epsilon|+\gamma+ \\
&
S\left(\epsilon p_0\tan\frac{\alpha'}{2}-\epsilon p_0\tan\frac{\alpha}{2}\right)\rbrace.
\end{split}
\end{equation}

With a bit of trigonometry we can separate integral operator into a circulant ($SO(2)$ symmetric part) and the rest: 
\begin{align}
-\frac{\pi p_0}{me^2}\chi(\alpha)=&\int_{-\pi}^{\pi}d\alpha'\chi(\alpha')
\log\left|\sin\left(\frac{\alpha-\alpha'}{2}\right)\right| \label{fe1}\\
+&\int_{-\pi}^{\pi}d\alpha'\chi(\alpha')
\log\left| \cos\left(\frac{\alpha'}{2}\right)\cos\left(\frac{\alpha}{2}\right)\right| \label{fe2}\\
+&\int_{-\pi}^{\pi}d\alpha'\chi(\alpha')\left\lbrace\log(p_0\epsilon e^\gamma) + S\left(\epsilon p_0\tan\frac{\alpha'}{2}-\epsilon p_0\tan\frac{\alpha}{2}\right)\right\rbrace \label{fe3}
\end{align}
The eigenfunctions of the integral operator (\ref{fe1}) follow from it's $SO(2)$ symmetry\footnote{See e.g. sec 3.5 of \cite{pip}.},  
\begin{equation}\label{so2sols}
\chi_n^\pm(\alpha)=e^{\pm in\alpha},
\end{equation}
with $n$ a positive integer and eigenvalues given by $p_0= me^2/n$. In this case appears double degeneracy of energy levels, which is from (\ref{so2sols}) interpreted as coming from a free semiclassical particle circling in positive or negative direction around "hypercircle". 

From here easily follow momentum space wavefunctions,
\begin{equation}\label{momsols}
\phi_n^\pm(p)=\frac{p_0^2}{p^2+p_0^2}e^{\pm2in\arctan \frac{p}{p_0}}=\frac{p_0^2}{p^2+p_0^2}\left(\frac{p_0+ip}{p_0-ip}\right)^{\pm n},
\end{equation}
same as in \cite{ny, dav}, as well as coordinate space wavefunctions\footnote{Explicit evaluation of Fourier transform in the appendix of \cite{pr}.},
\begin{equation}\label{coordsols}
\psi_n^\pm(x)=\Theta(\pm x)\frac{2p_0^{3/2}}{n}|x|e^{-|x|p_0}L_{n-1}^{(1)}(2|x|p_0)
\end{equation}
where $L_n^{(\alpha)}$ denotes a generalized Laguerre polynomial. These are the same as "Dirichlet" solutions in \cite{fis}.

Solutions (\ref{coordsols}) correspond to a particle constrained on a right/left coordinate space half-line. Particle confinement on a half-line is on one hand in agreement with  $SO(2)$ symmetry, since Laplace-Runge-Lenz vector is in 1d trivial, i.e. proportional to a unit vector $\mathbf{\hat x}/|x|$, and on the other it means breakdown of the parity symmetry, which is a one-dimensional remnant of the geometrical rotational symmetry of Kepler problem. In short,
$$
(\ref{fe1})\rightarrow SO(2)\rightarrow\ \text{double degeneracy}\nrightarrow SO(1)\ (\text{parity})=\text{confinment}$$

We next consider (\ref{fe1}) + (\ref{fe2}). Obviously (\ref{fe2}) explicitly brakes $SO(2)$ symmetry, but due to evenness of cosine function we can again immediately write eigenfunctions of the Hamiltonian operator, 

\begin{equation}\label{parsols}
\chi_n(\alpha)=\sqrt 2\sin(n\alpha),
\end{equation}
with $n$ a positive integer and with eigenvalues given by $p_0= me^2/n$. The momentum/coordinate space solutions are simply antisymmetric linear combinations of (\ref{momsols})/(\ref{coordsols}).  In this case parity is a symmetry of the problem, and with $SO(2)$ symmetry disappear also double degeneracy of energy levels, as well as particle confinement, and the solutions agree with the ones reported in \cite{gor, pr}. In short,  
$$
(\ref{fe1}) + (\ref{fe2})\nrightarrow SO(2)\nrightarrow\ \text{double degeneracy}\rightarrow SO(1)\ (\text{parity})=\text{barrier penetrability}$$

Finally we consider the full equation including the regularization dependent part, (\ref{fe1}) + (\ref{fe2})+(\ref{fe3}). The first term in (\ref{fe3}) is finite for all finite $\epsilon$, so it vanishes, for both (\ref{so2sols}) and (\ref{parsols}), even in the $\epsilon\to 0$ limit. The second term in (\ref{fe3}) explicitly breaks both symmetries for some finite $\epsilon$, and the solutions would be more involved then (\ref{so2sols}) and (\ref{parsols}), meaning also the first term in (\ref{fe3}) might have a nonvanishing contribution. In the $\epsilon\to0$ limit, the sum $S$ vanishes for finite momenta, but can have a nonvanishing (finite or infinite) value in the limit $p\to\infty$, depending on the definition of the double limit
$$\lim\limits_{\substack{\epsilon\to0 \\ p\to\infty}}\epsilon p.$$
Since this limit can be defined arbitrarily, there are infinite possibilities for defining (\ref{fe3}) in the $\epsilon \to0$ limit. For example, if one takes this limit to vanish, then (\ref{fe3}) adds nothing to the equation in the $\epsilon\to0$ limit, and the solutions are (\ref{parsols}). One can also ask if it is possible to define this limit in such a way that (\ref{fe3}) completely cancels (\ref{fe2}), bringing back $SO(2)$ symmetry. 

This is where the benefit of our approach becomes apparent. Regardless of the way one defines the above double limit, from eqs (\ref{fe1})-(\ref{fe3}) is evident that there is no way  to keep both parity and $SO(2)$ symmetries of the problem. Whichever regularization one uses, it tampers either (\ref{fe1}) and breaks $SO(2)$ symmetry, (\ref{fe2}) and breaks parity, or both.

There is a nice geometrical picture of the symmetry breakdown at one dimension. A $d$ dimensional Kepler problem can be embedded into $d+1$ dimension, meaning it inherits subgroups of $d+1$'s symmetry groups. This is simply achieved e.g. by constraining coordinate polar angle of the $d+1$ dimensional problem to $\pi/2$. For instance, starting with a 3d problem and constraining it to $\theta=\pi/2$ ($z=0$), one gets a 2d Kepler problem with both geometrical and dynamical symmetry. In passing to 1d from 2d, choosing a polar angle $\pi/2$ may define an $SO(2)$ subgroup of the $SO(3)$ dynamical symmetry, but it also automatically breaks parity by choosing only a half of coordinate line. In other words, for $d\ge2$ breaking of geometrical rotational symmetry implies breakdown of dynamical rotational symmetry as well, though the subgroup of the dynamical symmetry that rotates around the unphysical axes remains intact; when passing to a single dimension, all that remains of the dynamical symmetry are rotations around unphysical axes, which are independent of the geometrical rotational symmetry.

Finally, we note that the parity symmetry of a one-dimensional problem need not be considered a remnant of a geometrical rotational symmetry of higher dimensional problem upon it's reduction to a single dimension - it is possible to start with a higher dimensional problem without rotational symmetry, but possessing a reflection symmetry around $x=0$ and reduce it to a single dimension. This shows that when analysing a one-dimensional problem it is important to consider the context of the physical appearance of the problem from higher dimension.

\section{Conclusion}
We were able to analyse symmetry aspects of the 1d Kepler problem without specifying a singular point, which allowed us to demonstrate that a formulation that would keep both of the symmetries that are the signatures of Kepler problem is not possible in one dimension. We've also classified the solutions existing in the literature according to the symmetry they respect, and provided explanation for both of the strange features of the problem - particle confinement and double degeneracy of energy levels - in terms of the dynamical symmetry of the Kepler problem.

While it may seem like a technical point, we maintain that this curious observation, that managed to remain unnoticed in the literature, should be of some formal and even  practical value. The existence of different solutions is canonically interpreted  as due to arbitrariness in the definitions of the boundary conditions at the singular point, with all self-adjoint extensions of Hamiltonian (\ref{coord}) given in \cite{fis}. We showed that it can also be interpreted as due to different ways of breaking the symmetries of  Kepler problem upon it's reduction to a single dimension. Even in practical situations with real physical systems, it is usually very important how an effective 1d system is realised from higher dimension. To be able to classify a concrete problem according to it's symmetry as well, in addition to the self-adjoint extension to which it belongs, may show useful.

Finally, we have deliberately avoided the question of physicality of the solutions, which was, together with mathematical arbitrariness, the main fuel for such a scale of scientific research. In particular, one can not ignore the physical appeal of the solutions (\ref{parsols}), for they take away strange energy degeneracy and particle confinement, and it is understandable that in so many papers these were advocated. However, it is difficult to rely on physical argumentation without explicitly defining how one constrains a system to a single dimension or taking into account relativistic effects\footnote{As noted in \cite{gor}} or for instance radiation of an electron. And from mathematical point of view there is nothing wrong with the solutions (\ref{so2sols}) - they're eigenstates of a self-adjoint Hamiltonian \cite{fis}, which is from axiomatic point of view the only requirement. We plan to investigate this matter further in the future.

\section*{Acknowledgment}
I'm very grateful to my mentor S. Mignemi for his patience, encouragement and for many useful discussions without which this paper would not be possible.

\theendnotes
\vspace{10pt}
\noindent
E-mail:\textit{bivetic@yahoo.com}

\end{document}